\documentclass[onecolumn,11pt]{article}
\usepackage[top=1in, bottom=1in, left=1in, right=1in]{geometry}
\usepackage{amsmath,amsfonts,amscd,amssymb}
\usepackage{graphicx}
\usepackage{epstopdf}
\usepackage{overpic}
\usepackage{cancel}
\usepackage{rotating}
\usepackage{url}
\usepackage{caption}
\usepackage{color}
\usepackage{rotating}
\usepackage{multirow}
\usepackage{wrapfig}
\usepackage{mathtools}
\usepackage{subeqnarray}
\usepackage{setspace}
\usepackage{palatino} 
\usepackage[numbers,sort&compress]{natbib}
\usepackage{framed}
\usepackage{xcolor}
\usepackage{breqn}
\usepackage{enumitem}
\usepackage{mathtools}
\usepackage{pifont}
\newcommand{\cmark}{\ding{51}}
\newcommand{\xmark}{\ding{55}}

\usepackage[bottom,flushmargin,hang,multiple]{footmisc}
\usepackage{lipsum}
\newcommand\blfootnote[1]{%
  \begingroup
  \renewcommand\thefootnote{}\footnote{#1}%
  \addtocounter{footnote}{-1}%
  \endgroup
}

\newcommand{\bb}{\boldsymbol{b}}

\newcommand{\bx}{\boldsymbol{x}}

\newcommand{\bA}{\boldsymbol{A}}

\newcommand{\bP}{\boldsymbol{P}}

\newcommand{\bX}{\boldsymbol{X}}

\newcommand{\bPsi}{\boldsymbol{\Psi}}
\newcommand{\bphi}{\boldsymbol{\phi}}
\newcommand{\bPhi}{\boldsymbol{\Phi}}

\newcommand{\bLambda}{\boldsymbol{\Lambda}}
\newcommand{\bOmega}{\boldsymbol{\Omega}}

\DeclareMathOperator*{\argmin}{arg\rm{}min}

\definecolor{header1}{cmyk}{0,0,0,1}

\title{\LARGE{\vspace{-.55in}\textbf{Phase-consistent dynamic mode decomposition\\ from multiple overlapping spatial domains}}\vspace{-.175in}}
\title{\vspace{-.55in}{\fontsize{16}{16}\selectfont \textbf{Phase-consistent dynamic mode decomposition\\ from multiple overlapping spatial domains}}\vspace{-.15in}}

\author{\normalsize{Aditya G.~Nair$^{1*}$, Benjamin Strom$^{1}$, Bingni W. Brunton$^2$ and Steven L. Brunton$^1$}\\
\footnotesize{$^1$ Department of Mechanical Engineering, University of Washington, Seattle, WA 98195}\\
\footnotesize{$^2$ Department of Biology, University of Washington, Seattle, WA 98195\vspace{-.2in}}\\
}
\date{}
\begin{document}
\maketitle

\blfootnote{$^*$ Corresponding author (agnair@uw.edu).}
\vspace{-.2in}

\begin{abstract}
Dynamic mode decomposition (DMD) provides a principled approach to extract physically interpretable spatial modes from time-resolved flow field data, along with a linear model for how the amplitudes of these modes evolve in time.  
Recently, DMD has been extended to work with more realistic data that is under-resolved either in time or space, or with data collected in the same spatial domain over multiple independent time windows. 
In this work, we develop an extension to DMD to synthesize globally consistent modes from velocity fields collected independently in multiple partially overlapping spatial domains.
We propose a tractable optimization to identify modes that span multiple windows and align their phases to be consistent in the overlapping regions.  
First, we demonstrate this approach on data from direct numerical simulation, where it is possible to split the data into overlapping domains and benchmark against ground-truth modes.  
We consider the laminar flow past a cylinder as an example with distinct frequencies, along with the spatially developing mixing layer, which exhibits a frequency spectrum that evolves continuously as the measurement window moves downstream.  
Next, we analyze experimental velocity fields from PIV in six overlapping domains in the wake of a cross-flow turbine.  
On the numerical examples, we demonstrate the robustness of this approach to increasing measurement noise and decreasing size of the overlap regions. 
In all cases, it is possible to obtain a phase-aligned, composite reconstruction of the full time-resolved flow field from the phase-consistent modes over the entire domain.  
\end{abstract}

\section{Introduction}
Dynamic mode decomposition (DMD) has emerged as a leading algorithm to extract spatiotemporal coherent structures from high-dimensional time-series data of a fluid flow~\citep{Schmid:JFM10,Kutz2016book}.  
As with other modal decompositions~\citep{Taira2017aiaa,taira2019modal}, DMD relies on the fact that high-dimensional representations of a fluid often evolve on a low-dimensional attractor defined by coherent structures~\citep{berkooz1993proper,HLBR_turb}. 
The existence of these dominant flow patterns enables dimensionality reduction and the subsequent tasks of reduced-order modeling and flow control~\citep{Noack2011book,Benner2015siamreview,Brunton2015amr,Rowley2017arfm,Brunton2020arfm}.
More fundamentally, modal decomposition techniques provide insight into the underlying flow physics and nonlinear interaction mechanisms that drive flows~\citep{Lorenz1963jas,Noack2003jfm,schmid2012stability,Brunton2016pnas,Loiseau2017jfm,Taira2017aiaa,taira2019modal}.
%

Many factors have contributed to the widespread adoption of DMD. 
DMD is a data-driven and equation-free method that applies equally well to data from experiments or simulations. 
The DMD algorithm~\citep{Schmid:JFM10,Tu2014jcd} is typically based on the proper orthogonal decomposition (POD)~\citep{berkooz1993proper,HLBR_turb,Towne2018jfm} for dimensionality reduction, which extracts coherent structures hierarchically based on their ability to capture the most energy or variance in a flow. 
DMD has the additional constraint that each spatial mode must have the same linear behavior in time (i.e., oscillations at a given frequency, along with exponential growth or decay). 
Thus, DMD provides a reduced modal expansion, as well as a linear model for the evolution of the amplitudes of these modes in time.  
Although DMD results in linear models, it has been rigorously connected to nonlinear dynamical systems via the Koopman operator~\citep{Rowley:JFM09,mezic2013analysis,Kutz2016book}.  
Since its introduction by~\citet{Schmid:JFM10}, DMD has been applied to a variety of systems in fluid mechanics~\citep{schmid2011applications,lusseyran2011flow,schmid2012decomposition,semeraro2012analysis,seena2011dynamic,basley2013space} and more broadly, in fields as diverse as epidemiology~\cite{Proctor2015ih}, neuroscience~\cite{Brunton2014jnm}, robotics~\cite{Berger2014ieee}, and plasma physics~\cite{Taylor2018rsi,Kaptanoglu2019arxiv}; see~\cite{Kutz2016book} for a more complete list of examples with references.

Recently, a number of powerful extensions have been developed to make DMD applicable to more realistic data. 
The original DMD algorithm (introduced in \S\ref{subsec:DMD}) works well for time-resolved flow field measurements from a single time series with relatively little measurement noise.  
Although DMD has been shown to be sensitive to measurement noise~\cite{Bagheri2013jfm,Bagheri2014pof}, there are several algorithms to de-bias DMD results based on noisy~\citep{Dawson2016ef,Hemati2017tcfd,askham2018variable} or corrupt~\cite{Scherl2019arxiv} data. 
In addition, there are several extensions based on compressed sensing to extend DMD to data that is under-resolved in either time~\citep{tu2014eif} or space~\citep{Brunton2015jcd,Gueniat2015pof}.  
\citet{Tu2014jcd} further showed that it is possible to concatenate multiple time-series from independent simulations or experiments, with little modification to the DMD algorithm.  

\begin{figure}
  \centerline{\includegraphics[width=0.9\textwidth]{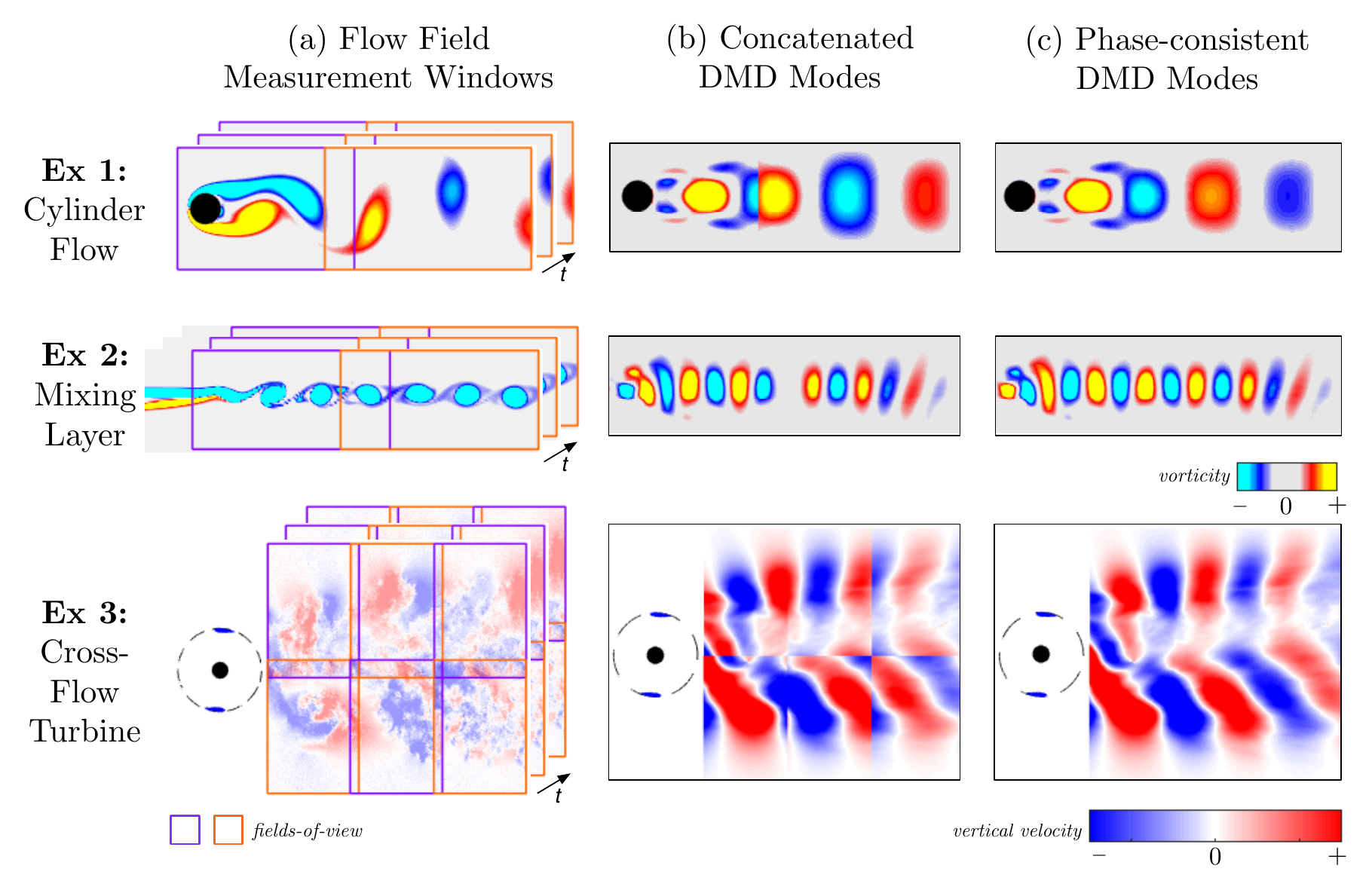}}
    \vspace{-.1in}
  \caption{Comparison between concatenated modes and the proposed phase-consistent DMD modes. For (a) data collected in multiple flow-field measurement windows, comparison between (b) concatenated DMD modes and (c) phase-consistent DMD modes. The examples considered include numerical simulations of flow over a cylinder and laminar mixing layer flow and PIV of a flow across a cross-flow turbine. The purple and orange boxes indicate the fields-of-view. } 
    \vspace{-.1in}
\label{fig1}
\end{figure}

In this work, we develop an extension to the DMD algorithm to synthesize velocity fields from multiple partially overlapping spatial domains into globally consistent modes.
Data from overlapping spatial domains is common in particle image velocimetry (PIV)~\citep{adrian2011cup},  where there is often a compromise between spatial resolution and the size of the measurement region. 
Building on the invariance of DMD to unitary transformations~\citep{Brunton2015jcd,Brunton2019book}, we develop a tractable optimization to align the phases of modes extracted in overlapping regions.  
In fact, the present work extends these previous invariance results, further describing how unitary transformations affect the DMD mode amplitudes.  
After developing the theoretical foundations for this phase-aligned DMD algorithm (discussed in \S \ref{subsec:paDMD}), we demonstrate it on several data sets from simulations and experiments, shown in Fig.~\ref{fig1}.  
We begin by validating this approach against ground truth modes obtained from direct numerical simulation, where it is possible to split the data into overlapping domains.  
%
The first numerical example is the laminar flow past a cylinder with distinct DMD frequencies (discussed in \S\ref{subsec:cylinder}), and the second numerical example considers the spatially developing mixing layer (discussed in \S\ref{subsec:mixing_layer}), with a continuously evolving frequency spectrum.  
The snapshots in these examples are collected in two overlapping domains at different times to mimic experimental data collection. 
The global modes and reconstructed snapshots synthesized over the entire domain are compared against ground truth modes.
We also systematically test the robustness of this method to measurement noise and the amount of overlap between domains. 
Finally, we demonstrate the utility of the proposed method on experimental velocity fields from PIV in six overlapping domains in the wake of a cross-flow turbine in \S\ref{subsec:cfturbine}.  
Concluding remarks are offered in \S\ref{conc}.

\section{Dynamic mode decomposition (DMD)}
\label{subsec:DMD}
The DMD algorithm decomposes high-dimensional fluid flow data into a set of spatial modes that evolve linearly in time~\citep{Kutz2016book}.  
In particular, consider a time-series of spatial flow field snapshots $\bx_k \triangleq\bx(t_k) \in \mathbb{R}^n$ (e.g., velocity, pressure, etc.~at $n$ spatial locations arranged in a column vector) sampled at times $t_k=k\Delta t$ for $k=0,1,2,...,m$; typically $n\gg m$.    
The DMD algorithm computes the leading eigendecomposition of a best-fit linear operator $\boldsymbol{A}$ so that 
\begin{equation}
\bx_{k+1} \approx \boldsymbol{A}\bx_k.  
\end{equation}
The eigenvectors $\boldsymbol{\phi}$ of $\bA$ are \emph{modes}, having the shape of a spatial flow field, $\boldsymbol{\phi}\in\mathbb{R}^n$, and their time dynamics are determined by the corresponding eigenvalues.  

The first step of the DMD algorithm is to arrange the time series data into matrices as 
\begin {equation}
{\boldsymbol{X}=  \begin{bmatrix} | & | & | & | \\ \bx_0 & \bx_1 & \cdots & \bx_{m-1} \\ | & | & | & |\end{bmatrix}},~~~~{\boldsymbol{X}^\prime =  \begin{bmatrix} | & | & | & | \\ \bx_1 & \bx_2 & \cdots & \bx_{m} \\ | & | & | & | \end{bmatrix}}
\end{equation} 
The best-fit linear operator $\boldsymbol{A}$ relates the two data matrices as 
\begin {equation}
{\boldsymbol{X}^\prime \approx \boldsymbol{A}\boldsymbol{X}},
\end{equation} 
and may be solved for via least-squares regression:
\begin{equation}
\boldsymbol{A} = \argmin_{\boldsymbol{A}}\|\boldsymbol{X}'-\boldsymbol{A}\boldsymbol{X}\|_F = \boldsymbol{X}'\boldsymbol{X}^{\dagger}
\end{equation}
where $\|\cdot\|_F$ is the Frobenius norm and $\boldsymbol{X}^{\dagger}$ is the pseudo-inverse, which is computed via  singular value decomposition of $\boldsymbol{X}$:
\begin {equation}
\boldsymbol{X}=\boldsymbol{U\Sigma V}^T \quad\Longrightarrow\quad \boldsymbol{X}^{\dagger}=\boldsymbol{V} \boldsymbol{\Sigma}^{-1} \boldsymbol{U}^T. 
\end{equation}
The columns of $\boldsymbol{U}$ are POD modes arranged hierarchically by how much of the variance of the matrix $\boldsymbol{X}$ they contain, quantified by the singular values $\sigma_j$ from the non-negative, diagonal matrix $\boldsymbol{\Sigma}$. 
The matrices $\boldsymbol{U}$ and $\boldsymbol{V}$ are unitary, so that $\boldsymbol{U}^T\boldsymbol{U} = \boldsymbol{I}$ and $\boldsymbol{V}^T\boldsymbol{V} = \boldsymbol{I}$, where $\boldsymbol{I}$ is the identity matrix.  
Thus, the matrix $\boldsymbol{A}$ may be solved for as  
\begin {equation}
\boldsymbol{A} = \boldsymbol{X}^\prime \boldsymbol{V \Sigma}^{-1}\boldsymbol{U}^T.
\end{equation} 

For high-dimensional systems, the matrix $\boldsymbol{A}$ may be exceeding large, so that it is unreasonable to directly represent, let alone compute, its eigenvalues and eigenvectors.  
Instead, a rank $r$ approximation of $\boldsymbol{X}$ is used, $\boldsymbol{X}\approx\boldsymbol{U}_r\boldsymbol{\Sigma}_r \boldsymbol{V}_r^T$, where $\boldsymbol{U}_r$ and $\boldsymbol{V}_r$ correspond to the first $r$ columns and $\boldsymbol{\Sigma}_r$ corresponds to the first $r\times r$ block.  
The matrix $\boldsymbol{A}$ is then projected onto the leading $r\leq m\ll n$ POD modes: 
\begin{equation}
{\boldsymbol{A}_r} = \boldsymbol{U}_r^T\boldsymbol{A}\boldsymbol{U}_r = \boldsymbol{U}_r^T\boldsymbol{X}^\prime \boldsymbol{V}_r \boldsymbol{\Sigma}_r^{-1}.
\end{equation}
It is computationally tractable to compute the eigendecomposition 
\begin{equation}
{\boldsymbol{A}_r}\boldsymbol{W} = \boldsymbol{W}\boldsymbol{\Lambda}
\end{equation}
of the $r\times r$ matrix ${\boldsymbol{A}_r}$, and the eigenvalues of ${\boldsymbol{A}_r}$ are also eigenvalues of ${\boldsymbol{A}}$.  
The eigenvectors $\boldsymbol{\Phi}$ of the full matrix $\boldsymbol{A}$ may then be computed from the reduced eigenvectors $\boldsymbol{W}$ via 
\begin{equation}
{\boldsymbol{\Phi} = \boldsymbol{X}^\prime \boldsymbol{V}\boldsymbol{\Sigma}^{-1}\boldsymbol{W}}.
\end{equation}
This formulation is known as \emph{exact DMD}~\citep{Tu2014jcd}, and the original DMD paper~\cite{Schmid:JFM10} computed \emph{projected DMD} modes as ${\boldsymbol{\Phi} = \boldsymbol{U}\boldsymbol{W}}$. 
To be consistent in our results, we have used exact DMD for all the analysis in this paper.

The diagonal matrix $\bLambda$ contains DMD eigenvalues, which determine the linear behavior of the corresponding modes, given by the columns of $\bPhi$.  
Given DMD modes and eigenvalues, it is possible to represent the data as:
\begin{align}
\bx_k \approx \sum_{j=1}^r \bphi_j\lambda_j^k b_j = \bPhi \bLambda^k \bb. 
\end{align}
The vector $\bb$ contains the complex-valued mode amplitudes, which are computed from 
\begin{equation}
\bb = \bPhi^{\dagger}\bx_0,
\end{equation}
where $\dagger$ denotes the pseudo-inverse.
The sparsity-promoting DMD~\cite{jovanovic2014sparsity} provides an alternative, computing $\bb$ with as many zero entries as possible, extracting dominant modes.  
There is also a noise-robust DMD alternative, called optimized DMD~\citep{askham2018variable}. 

Equivalently, it is possible to represent the DMD in continuous time.  
The discrete-time eigenvalues ${\lambda}$ may be converted to continuous time via ${\omega} = \log({\lambda})/\Delta t$; the real part of the eigenvalue is the growth/decay rate  and the imaginary part determines the oscillation frequency ${\omega}$. 
The DMD reconstruction then becomes
\begin{align}
\bx(t) \approx  \bPhi \exp(\bOmega \, t)\bb. \label{eq2}
\end{align}

For oscillatory flows in the absence of measurement noise, DMD modes and eigenvalues come in complex conjugate pairs that are on the imaginary axis in continuous-time and on the unit circle in discrete-time.  
Gaussian measurement noise appears as artificial damping in these eigenvalues~\cite{Bagheri2014pof}.  
The DMD modes $\bPhi$ will likewise come in complex conjugate pairs, and will thus have a magnitude and a phase.  
In fact, any arbitrary phase shift $\bPsi$ may be factored out of the DMD modes and incorporated into the exponential time dynamics, or the mode amplitudes, in Eq.~\eqref{eq2}:
\begin{align}
\tilde\bPhi = \bPhi \exp(i\bPsi)\quad\Longrightarrow\quad \bx(t) = \tilde\bPhi \exp(\bOmega t - i\bPsi)\bb = \tilde\bPhi \exp(\bOmega t) \tilde\bb,
\end{align}
where $\tilde\bb = \exp(-i\bPsi)\bb$.  Thus, the phase shift can be transferred from the modes $\bPhi$ to the amplitudes $\bb$, and vice versa.  
Identifying the phase shift $\bPsi$ to make DMD consistent in the overlap regions from multiple independent experiments will be the subject of the next section.  

\section{Phase-consistent DMD}
\label{subsec:paDMD}
In experiments, it is often difficult to obtain time-resolved measurements with high spatial resolution over a large domain of interest. 
Instead, it is common to collect data from multiple overlapping spatial domains $\mathcal{D}^{(j)}$ from $j = 1,2,\cdots ,p$, for example using PIV. 
While each of these datasets may be time-resolved, they are collected at different initial times starting from different initial flow conditions. 
If the flow is stationary and has dominant periodic or quasi-periodic behavior, then it is possible to align the phases of the DMD modes in each of the overlapping windows to approximate the global DMD modes, and subsequently the phase-consistent flow field, over the entire domain $\mathcal{D}^{\cup} \triangleq \bigcup_{j=1}^p \mathcal{D}^{(j)}$.  
However, for spatially developing flows, non-stationary flows, aperiodic flows, and more generally for turbulent flows, this approach may break down, as we will explore in the examples.

If data $\boldsymbol{X}^\cup$ is available on the entire domain $\mathcal{D}^\cup$, then it is possible to compute the global DMD modes $\boldsymbol{\Phi}^\cup$ directly.  
Instead, we have snapshot data $\{\boldsymbol{X}^{(j)},\boldsymbol{X}^{'(j)}\}_{j=1}^p$ collected in domains $\mathcal{D}^{(j)}$: 
\begin{equation}
{\boldsymbol{X}^{(j)} =  \begin{bmatrix} | & | & | & | \\ \bx_{m_j}^{(j)} & \bx_{m_j+1}^{(j)} & \cdots & \bx_{m_j+m-1}^{(j)} \\ | & | & | & | \end{bmatrix}},~~~~~{\boldsymbol{X}^{'(j)} =  \begin{bmatrix} | & | & | & | \\ \bx_{m_j+1}^{(j)} & \bx_{m_j+2}^{(j)} & \cdots & \bx_{m_j+m}^{(j)} \\ | & | & | & |\end{bmatrix}}. 
\label{eqIDMD}
\end{equation}
Here, all datasets start at different initial times $t_{m_j}$ corresponding to different initial conditions.  
For simplicity, all datasets also contain $m$ snapshots, although this may be relaxed.  
Given two spatial domains $\mathcal{D}^{(j)}$ and $\mathcal{D}^{(l)}$, we denote the overlap region as $\mathcal{D}^{(j,l)} \triangleq \mathcal{D}^{(j)}\cap \mathcal{D}^{(l)}$.
The snapshot from $\boldsymbol{X}^{(j)}$ at time $t_k$ restricted to the overlap region $\mathcal{D}^{(j,l)}$ will be denoted by $\bx_{k}^{(j)|(j,l)}$.

Our overall objective is to develop an algorithm that synthesizes data from multiple overlapping domains into globally consistent DMD modes.  
However, if we naively compute DMD on data from each domain $\mathcal{D}^{(j)}$ in isolation, then the phases of the modes will not agree in the overlap region, i.e., $\bPhi^{(j)|(j,l)}$ and $\bPhi^{(l)|(j,l)}$ are not the same for two domains $\mathcal{D}^{(j)}$ and $\mathcal{D}^{(l)}$.  
Thus, it is necessary to adjust the phases of the modes to be consistent in the overlap region. 
We denote the phase-consistent modes in domain $\mathcal{D}^{(j)}$ as 
\begin{equation}
\tilde{\bPhi}^{(j)} = \bPhi^{(j)}\exp\left(i{\bPsi}^{(j)}\right),
\end{equation}
where ${\bPsi}^{(j)}$ is a diagonal matrix containing the phase shifts for all modes.   

\begin{figure}
\vspace{-.2in}
  \centerline{\includegraphics[width=0.95\textwidth]{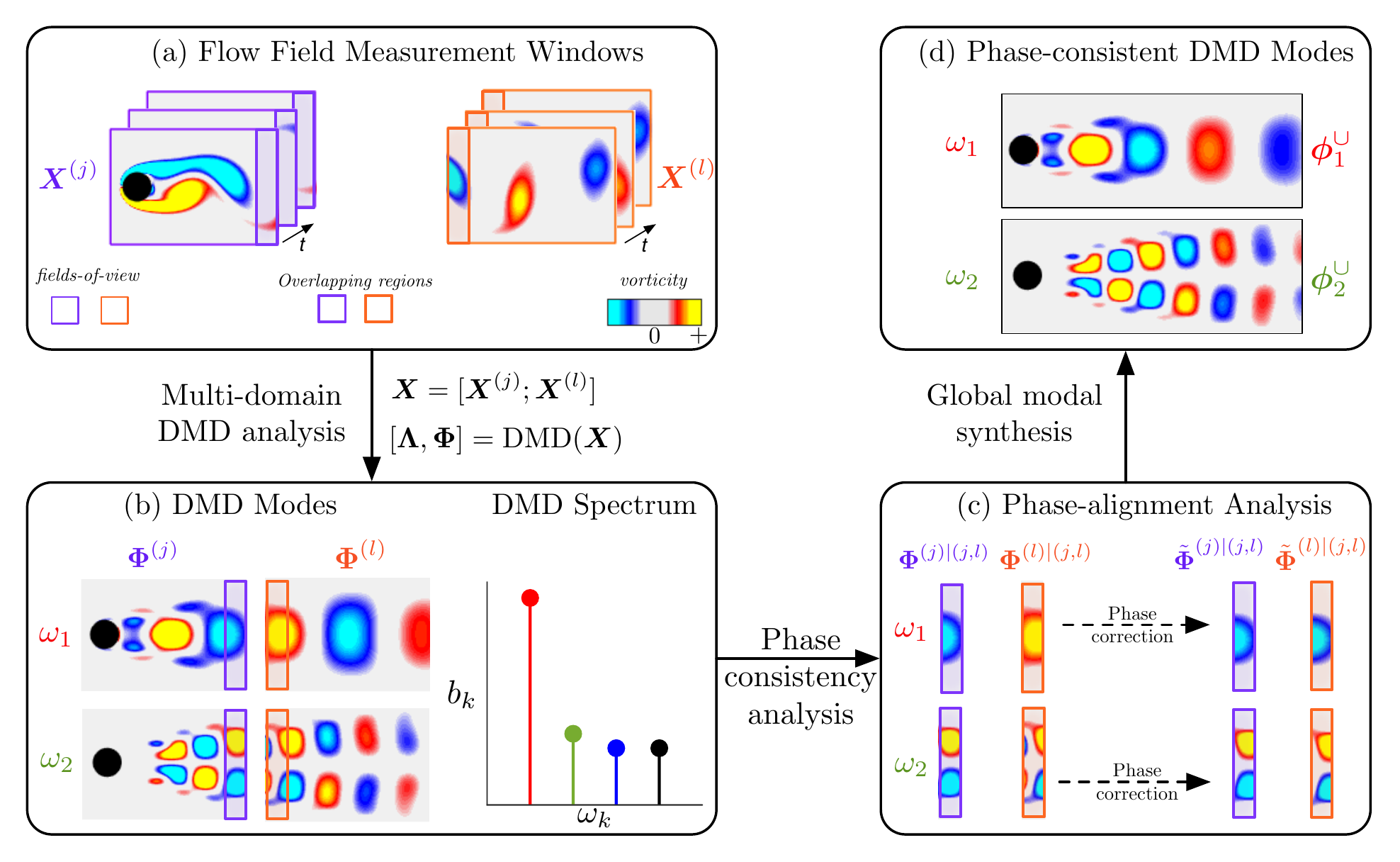}}
    \vspace{-.1in}
  \caption{Overview of the modal phase-alignment approach for flow over a cylinder at $Re = 100$: (a) Flowfield snapshots collected over multiple fields-of-view with an overlapping region, (b) DMD modes and spectrum from DMD analysis performed over all the data collected, (c) Phase-alignment using the spatial modal basis in the overlapping regions, (d) Reconstruction of the phase-consistent modes.}
    \vspace{-.15in}
\label{fig2}
\end{figure}

The overall analysis structure (see Fig.~\ref{fig2} for a particular flow example) is summarized as  
 \begin{equation}
\left\{\boldsymbol{X}^{(j)},\boldsymbol{X}^{'(j)}\right\}_{j=1}^p \xrightarrow[~\text{  analysis  }~]{\text{1. DMD}}  \left\{\bPhi^{(j)}\right\}_{j=1}^p \xrightarrow[~\text{  consistency  }~]{\text{2. Phase}} \left\{\tilde{\bPhi}^{(j)}  \right\}_{j=1}^p \xrightarrow[~\text{  synthesis  }~]{\text{3. Global}}\tilde{\bPhi}^{\cup}.
 \end{equation}  

There are choices for each of the three main steps in this procedure:  
\begin{enumerate}[noitemsep,topsep=0pt]
\item Perform DMD on data $\left\{\boldsymbol{X}^{(j)},\boldsymbol{X}^{'(j)}\right\}_{j=1}^p$ from multiple domains to obtain modes $\left\{\bPhi^{(j)}\right\}_{j=1}^p$. This will be discussed in \S \ref{subsec:mdDMD}.  Although the most natural approach is to compute DMD separately on each domain, we will show that computing DMD on a concatenated data set makes it easier to identify modes of the same frequency across domains.  
\item Perform phase-consistency analysis to obtain phase-consistent modes $\left\{\tilde{\bPhi}^{(j)}\right\}_{j=1}^p$ using the modes obtained from step 1. This will be discussed in \S \ref{subsec:pcDMD}.  In particular, we will introduce a simple and illustrative approach for time-periodic flows over two overlapping domains, followed by a general optimization procedure for multiple overlapping domains.  
\item Global mode synthesis $\tilde{\bPhi}^\cup$ from phase-consistent modes in each overlapping domain.  This will be discussed in \S \ref{gms}. 
\end{enumerate}

\subsection{Multi-domain DMD}
\label{subsec:mdDMD}

We remark that the simplest choice is to perform DMD independently on the data $\{\boldsymbol{X}^{(j)},\boldsymbol{X}^{'(j)}\}$ from each domain in isolation.  
However, this approach will lead to a different set of DMD eigenvalues $\bOmega^{(j)}$ and mode amplitudes $\bb^{(j)}$ in each domain. 
Even for periodic flows, it is difficult to associate modes from different domains corresponding to the same frequency, since small amounts of measurement noise will cause the eigenvalues to differ between windows.  
Moreover, without noise, the amplitudes still vary considerably in different windows.  

Instead, we compute DMD on data concatenated from the $p$ overlapping spatial domains
\begin{equation}
\boldsymbol{X} =\begin{bmatrix} 
\boldsymbol{X}^{(1)} \\  
\boldsymbol{X}^{(2)} \\  
\vdots \\
\boldsymbol{X}^{(p)} 
\end{bmatrix}, ~~~~~
\boldsymbol{X}^\prime = \begin{bmatrix} 
\boldsymbol{X}^{\prime(1)} \\
\boldsymbol{X}^{\prime(2)} \\
\vdots \\
\boldsymbol{X}^{\prime(p)} 
\end{bmatrix}.
\label{eqCDMD}
\end{equation}
Performing DMD on this concatenated dataset results in concatenated DMD modes 
\begin{equation}
\bPhi = \begin{bmatrix}
\bPhi^{(1)}\\
\bPhi^{(2)}\\
\vdots \\
\bPhi^{(p)}
\end{bmatrix}.
\label{Phieq}
\end{equation}  
Unlike performing DMD on each domain in isolation, these concatenated modes all share the same DMD eigenvalues $\bOmega$ and amplitudes $\bb$, so that it is natural to associate modes from each domain corresponding to the same frequency.  
However, modes with oscillatory behavior (i.e., complex conjugate eigenvalues) will generally not agree on the overlap regions, because they will have a different phase of oscillation. 
This motivates the phase-consistency analysis in the next section.

\subsection{Phase-consistency analysis}
\label{subsec:pcDMD}

To align the phases of the concatenated DMD modes from multiple domains, we first present a simple illustrative approach suitable for strictly periodic flows with only two overlapping domains.
Building on the illustrative example, we then present a general approach suitable for complex flows with any number of overlapping domains. 

\subsubsection{Simple illustrative approach}

This illustrative approach assumes that the data is strictly periodic and there are only two overlapping domains: $\mathcal{D}^{(1)}$ and $\mathcal{D}^{(2)}$.
As the data in these overlapping regions are from different experimental runs, the modes extracted ($\bPhi^{(1)}$ and $\bPhi^{(2)}$) differ in the phase of oscillation. 
The goal is to deduce the relative phase shift $\tilde\bPsi$ between the modes, so that it is possible to synthesize phase-consistent DMD modes across the overlapping domains:    
\begin{align}
\tilde{\bPhi} = \begin{bmatrix}
{\bPhi}^{(1)}\\
{\bPhi}^{(2)} \exp(i\bPsi) 
\end{bmatrix}.
\label{bPhi}
\end{align}
Here $\tilde{\bPhi}$ denotes the matrix of concatenated, phase-consistent DMD modes.  Without loss of generality, we align the phase of the second domain to the first.

We now illustrate a simple approach to extract this relative phase shift ${\bPsi}$.
Once DMD is performed on the concatenated dataset, a flow field snapshot in domain $\mathcal{D}^{(j)}$ at a given time can be reconstructed from $\bPhi^{(j)}$, $\bOmega$, and $\bb$, as in Eq.~\eqref{eq2}.    

In the overlap region, we have data collected from two independent experiments that are initialized at different times and phases.  
Recall that the DMD eigenvalues $\bOmega$ and amplitudes $\bb$ are the same in both domains, and only the modes differ.  Further, if the flow is strictly periodic, then the modes from each domain will have the same magnitude and will only differ in the phase of oscillation.  
If we want to approximate the snapshots from the first experiment in the overlap region with modes from the second experiment, it is necessary to introduce the phase shift $\exp(i\bPsi)$:
\begin{subequations}
\begin{align}
\bx^{(1)|(1,2)}(t) & = \bPhi^{(1)|(1,2)}\exp(\bOmega t)\bb\\
& = \tilde{\bPhi}^{(2)|(1,2)}\exp(\bOmega t) \bb\\ 
& = {\bPhi}^{(2)|(1,2)}\exp(i\bPsi)\exp(\bOmega t) \bb.
\end{align}
\end{subequations}
Solving for the phase shift $\exp(i\bPsi)$ yields:
\begin{equation}
\exp(i\bPsi) \approx ({\bPhi}^{(2)|(1,2)})^{\dagger} {\bPhi}^{(1)|(1,2)}.
\end{equation}
Note that the phase-shift operator $ \exp(i\bPsi)$ is a unitary operator, and it was shown in \citet{Brunton2015jcd} that DMD modes are invariant to left and right unitary transformations of data.

Once the phase correction is obtained, it is possible to synthesize phase-consistent DMD modes across the overlapping domains using Eq.~(\ref{bPhi}); the phase-consistent modes may then be synthesized into modes across the entire domain $\tilde{\bPhi}^{\cup}$ as described in \S \ref{gms}. 
Although this approach is illustrative, it is not simple to perform this regression when there are more than two overlapping domains, motivating the general approach in the next section.  

\subsubsection{General approach}\label{Sec:GeneralPhaseDMD}

As discussed above, finding the relative phase shift between overlapping regions is key in synthesizing global DMD modes. 
When we have multiple overlapping domains, we need to compute the relative phase shift for the modes in each domain to achieve global phase consistency. 
Instead of computing the phase shifts for all pairs of domains, as in the previous section, we find the optimal phase shift for the modes in each domain simultaneously by setting up an optimization problem that minimizes the residual error among all overlap regions. 
We achieve this by aligning the phases of each DMD mode across all domains.  
The objective is to identify phase shifts $\bPsi^{(j)}$ for each domain $\mathcal{D}^{(j)}$ to make the phase of the modes consistent in the overlap regions:  
\begin{align}
\tilde{\bPhi} = \begin{bmatrix}
{\bPhi}^{(1)}\exp(i\bPsi^{(1)})\\
{\bPhi}^{(2)}\exp(i\bPsi^{(2)})\\
\vdots \\
{\bPhi}^{(p)}\exp(i\bPsi^{(p)})\\
\end{bmatrix}.
\label{phiequation}
\end{align}
In particular, we minimize the $L_2$-error of the difference of the phase-consistent modes in the overlap regions according to the following optimization problem:
\begin{align}
\left\{\bPsi^{(j)}\right\}_{j=1}^p = \argmin_{\left\{\bPsi^{(j)}\right\}_{j=1}^p} \sum\limits_{j\neq l}\left\|\bPhi^{(j)|(j,l)}\exp(i\bPsi^{(j)}) - \bPhi^{(l)|(j,l)}\exp(i\bPsi^{(l)})\right\|_2^2.
\end{align}
Note that in practice, this optimization problem may be solved for each DMD mode separately. Solving for the optimal phase shifts for each mode separately in all fields-of-view eases the computational load. 
The optimization is solved using a simple Nelder-Mead simplex algorithm~\citep{lagarias1998convergence}. 

\subsection{Global mode synthesis}
\label{gms}

To synthesize global DMD modes over the entire domain $\tilde{\bPhi}^{\cup}$, we perform a weighted average of the phase-consistent modal values over the overlapping regions.
If the phase-consistency analysis is performed accurately, the modal values in the overlapping domains are quite similar. 
The weighted average is obtained by a sliding neighborhood operation \cite{bucy1970linear} such that region closest to one of the overlapping domain gets weighted more and the region in the center of the overlapping region gets weighted equally. 
In the non-overlapping region, the modal values corresponding to that region is used for global mode synthesis.

\subsection{Relationship to unitary transformations}

The results in this paper are closely related to theoretical results from Brunton et al.~\cite{Brunton2015jcd}, which describes how unitary transformations on the data affect the resulting DMD.  
In particular, it was shown that a right unitary transformation $\bP$, for example permuting the columns of $\bX$ and $\bX'$ to $\bX\bP$ and $\bX'\bP$, would leave the DMD modes and eigenvalues unchanged.  
However, this description was incomplete, and neglected to consider how the DMD amplitudes $\bb$ are affected by unitary transformations.  
It is now clear that although right unitary transformations do not change the DMD modes and eigenvalues, they do modify the mode amplitudes.  
For example, consider a transformation $\bP$ that shifts the \emph{periodic} data in $\bX$ and $\bX'$ by $k$ steps:
\begin{align*}
\begin{bmatrix} \bx_1 & \cdots &  \bx_k & \bx_{k+1} & \cdots & \bx_m\end{bmatrix} \bP = \begin{bmatrix} \bx_{1+k} & \cdots & \bx_m & \bx_1 & \cdots & \bx_{m+k}\end{bmatrix}. 
\end{align*}
The results of DMD on the shifted data are $\left\{\bPhi,\bLambda, \tilde{\bb} = \left(\bLambda^k\right)\bb\right\}$.
The phase shifts from Eq.~\eqref{phiequation} establish the unitary transformation that shifts the data in each window so the mode phases align.

\section{Results}
\label{res}
We now demonstrate the proposed phase alignment procedure on three example fluid flows of increasing complexity, shown in Fig.~\ref{fig1}. 
For the first two examples of flow past a cylinder and a laminar free-shear layer, data are collected in two domains from numerical simulations, where the ground truth is available for benchmarking.  
In the third example, PIV fields are collected for an experimental flow over a cross-flow turbine in six overlapping fields-of-view. 
Fig.~\ref{fig1}(a) shows the measurement windows used for the examples considered.
Without any phase correction, concatenated modes are clearly misaligned in the overlapping regions, as shown in Fig.~\ref{fig1}(b). 
After aligning the phases appropriately, globally phase-consistent modes are obtained, shown in Fig.~\ref{fig1}(c).
For the flows considered in this work, Table~\ref{table1} summarizes the nature of the flow, availability of ground truth, sampling time between snapshots ($\Delta t$), number of independent domains ($p$), number of snapshots considered in each domain ($m$), and  truncation rank ($r$) for DMD analysis.

\begin{table}[ht]
\vspace{-.05in}
\centering 
\caption{Parameters used for three example fluid flows. } 
\vspace{-.1in}
\begin{tabular}{c c c c c c c c} 
\hline\hline 
Case & Data Source &Nature & Ground truth & $\Delta t$ & $p$ & $m$ & $r$\\ [0.5ex] 
\hline 
Cylinder flow & Simulation& periodic & \cmark & 0.08 & 2 & 229 & 10\\ 
Mixing layer  & Simulation& aperiodic & \cmark & 0.002 & 2 & 600 & 6\\
Cross-flow turbine & Experiment& quasi-periodic & \xmark & 0.01 & 6 & 1000 & 14\\ [0.5ex] 
\hline 
\end{tabular}
\vspace{-.2in}\label{table1} 
\end{table}

\subsection{Laminar flow past a cylinder}
\label{subsec:cylinder}
We first demonstrate the approach for a canonical example of two-dimensional incompressible flow past a circular cylinder at Reynolds number $Re = 100$, based on the cylinder diameter $d$. 
Data is generated via direct numerical simulations (DNS) using the immersed boundary projection method \citep{Taira:JCP07,Colonius:CMAME08, kajishima2017numerical}. 
A multi-domain technique is used with an inner domain of $x/d \in [-1,29], y/d \in [-15,15]$ and a resolution of $600 \times 600$ grid points; $x$ is the streamwise direction and $y$ is the cross-stream direction. 
Numerical details can be found in \cite{nair2018networked}. 
The flow exhibits periodic vortex shedding in the wake with Strouhal number $St = 0.164$ and period $T = 1/St \approx 6.1$.

In this simulated flow, we have access to time-resolved measurements over the entire domain for a large number of vortex shedding periods.  
Thus, it is possible to artificially create two overlapping domains that are sampled at different initial times, and then compare the phase-reconstructed flow fields to the ground truth simulation over the full domain.  
The full and overlapping domain extents are summarized in Table \ref{table2}. 
For the phase-consistency analysis, vorticity snapshots $\boldsymbol{X}^{(1)}$ are collected in domain $\mathcal{D}^{(1)}$ and snapshots $\boldsymbol{X}^{(2)}$ are collected in $\mathcal{D}^{(2)}$. 
The snapshots in the two domains are collected at different initial times: 
In domain $\mathcal{D}^{(1)}$, the snapshots are collected  at times $t \in [0,3T]$, while in domain $\mathcal{D}^{(2)}$, the snapshots are collected at times $t \in [3.3T,6.3T]$. 

\begin{table}
\vspace{-.125in}
\centering 
\caption{Dimensions of overlapping domains for flow past a cylinder example. }
\vspace{-.1in}
\begin{tabular}{c c c c c c c c} 
\hline\hline 
Full domain & $x/d$ & $y/d$ & Overlap. domains & $x/d$ & $y/d$ & $(\%)$ overlap\\ [0.5ex] \hline 
& & & $\mathcal{D}^{(1)}$ & $[-1,5]$ & $[-2,2]$  &  \\[-1ex] 
\raisebox{1.5ex}{$\mathcal{D}$} & \raisebox{1.5ex}{$[-1,10]$} & \raisebox{1.5ex}{$[-2,2]$} & $\mathcal{D}^{(2)}$ & $[4,10]$ & $[-2,2]$  & \raisebox{1.5ex}{$16.67$}\\[0.5ex] 
\hline 
\end{tabular}
\label{table2} 
\vspace{-0.175in}
\end{table}

The overall phase alignment procedure for this example is shown in Fig.~\ref{fig2}. 
The multi-domain DMD analysis from \S \ref{subsec:mdDMD} is used to extract $\boldsymbol{\Phi}^{(1)}$ and $\boldsymbol{\Phi}^{(2)}$. 
The DMD modes, frequencies, and amplitudes, are shown in Fig.~\ref{fig2}(b). 
The first mode pair corresponds to the vortex shedding frequency, and subsequent modes correspond to higher harmonics.  
There is a clear discrepancy in the phase of the modes in the overlap region.  
The phase shift required to align the modes is shown in Fig.~\ref{fig2}(c), and the globally consistent phase-shifted modes are shown in Fig.~\ref{fig2}(d).

\subsubsection{Effect of overlap window size and noise}

For experimental applications, it is important that this approach is robust to noise and various overlap sizes, and here we investigate how these affect the phase-consistent DMD analysis. 
The amount of overlap is computed as the ratio of the area of the overlap region to the area of the full domain.
Intuitively, small overlap and large noise will be more challenging than large overlap and small noise.  
To generate ground-truth results for comparison with our approach, we perform DMD on snapshots from the full domain $\mathcal{D}$, initialized at the same time as snapshots in $\mathcal{D}^{(1)}$.

\begin{figure}
  \centerline{\includegraphics[width=0.9\textwidth]{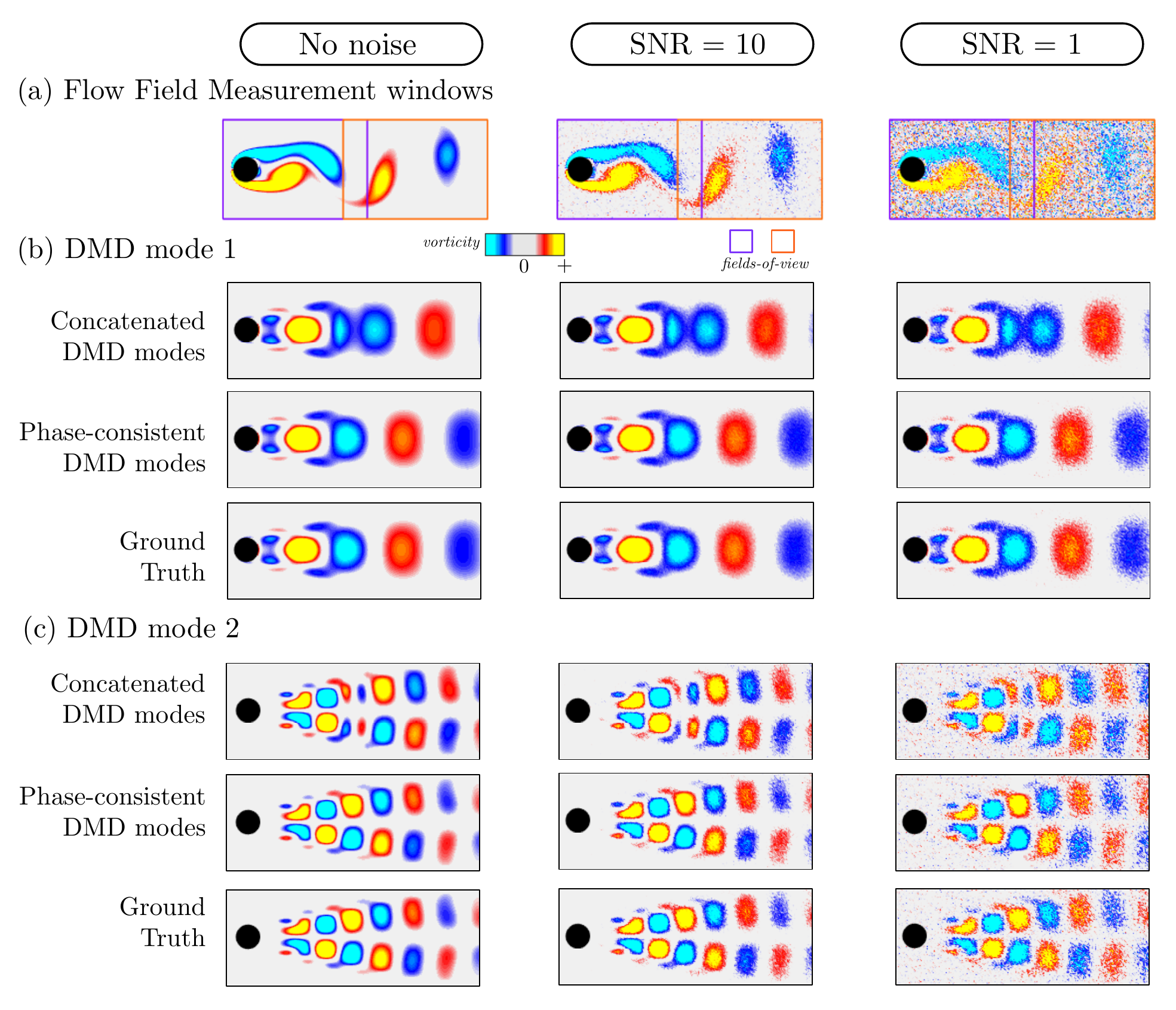}}
    \vspace{-.1in}
  \caption{Phase consistency analysis for flow over a cylinder at $Re = 100$: For (a) data collected in multiple flow-field measurement windows, comparison of concatenated modes, phase-consistent modes and ground truth for (b) DMD mode 1 and (c) DMD mode 2 for datasets with no added noise (left), datasets with SNR~$= 10$ (middle) and datasets with SNR~$= 1$ (right).}
    \vspace{-.15in}
\label{fig3}
\end{figure}

To test the robustness of the approach to noise, we add Gaussian noise to the vorticity snapshots, and explore a range of signal to noise ratios (SNRs).
The flow fields and modes are shown in Fig.~\ref{fig3}(a) for cases with no noise, low noise (SNR $= 10$), and high noise (SNR $= 1$). 
The phase-consistent modes ($\tilde{\bPhi}^\cup$) are compared with ground-truth modes in Fig.~\ref{fig3}(b) and (c), showing good agreement in all cases. 
We also compute the corresponding concatenated DMD modes (${\bPhi}^{\text{cat}}$), which are the raw output of the multi-domain DMD analysis from  \S \ref{subsec:mdDMD} before correcting the phases. 
The concatenated modes show significant error in the overlapping regions for all cases.

\begin{figure}
  \centerline{\includegraphics[width=0.9\textwidth]{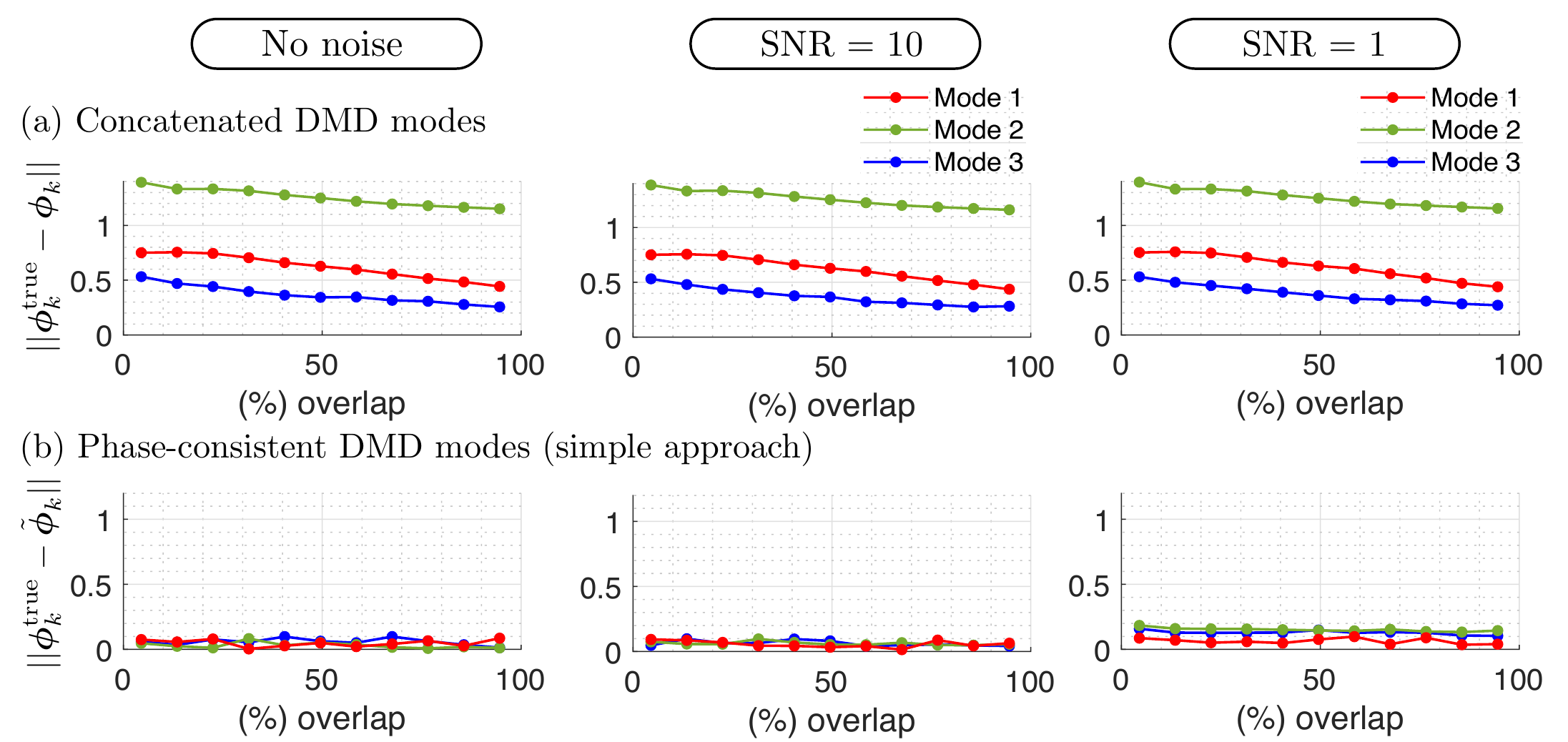}}
    \vspace{-.1in}
  \caption{Error in phase-consistency analysis for cylinder flow data with no noise, SNR $= 10$ and SNR $= 1$ : (a) Error in concatenated DMD modes, (b) Error in phase-consistent DMD modes. }
    \vspace{-.15in}
\label{fig4}
\end{figure}

\begin{figure}
  \centerline{\includegraphics[width=0.9\textwidth]{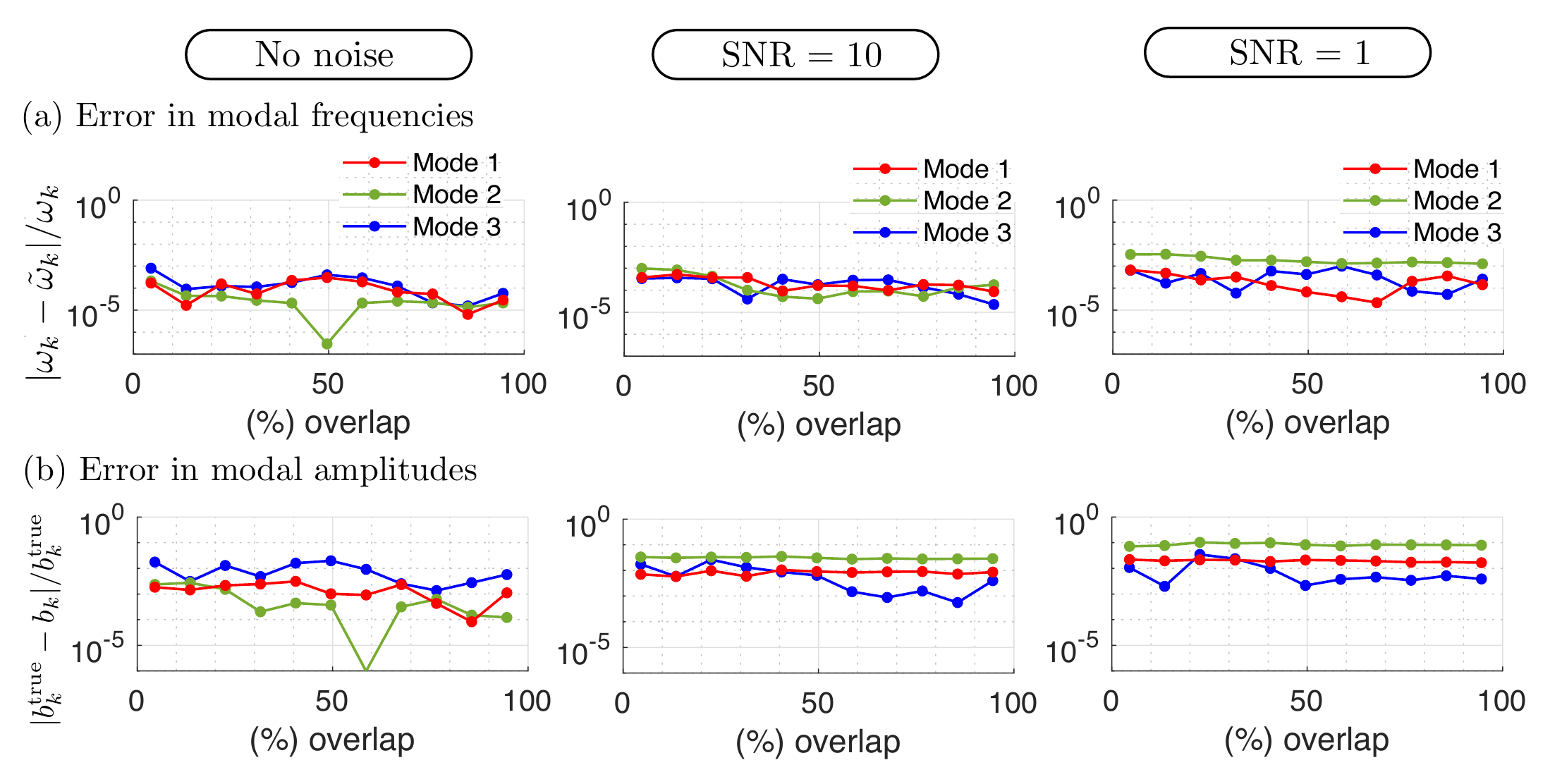}}
    \vspace{-.1in}
  \caption{Error in multi-domain DMD analysis for cylinder flow data with no noise, SNR $= 10$ and SNR $= 1$: (a) Error in modal frequencies and (b) Error in modal amplitudes vs. overlap size.}
    \vspace{-.15in}
\label{fig5}
\end{figure}

Figure~\ref{fig4} shows how noise and the overlap size affect the phase-shifted modes. 
The $L_2$ error between the concatenated modes (${\bPhi}^{\text{cat}}$) and ground truth modes (${\bPhi}^{\cup}$) is shown in Fig.~\ref{fig4}(a). 
Regardless of the amount of noise, there is substantial error. 
The error decreases slightly as the overlap region increases, although it is still significant.
The error between the phase-consistent DMD modes ($\tilde{\bPhi}^\cup$) and ground truth modes  (${\bPhi}^{\cup}$) is shown in Fig.~\ref{fig4}(b). 
The error in the phase-consistent modes is significantly lower than in the concatenated modes.  
The phase-consistent modes are quite robust to noise, only showing a slight increase in error with increasing noise.  
Similarly, the phase-consistent modes are extremely robust to variations in the overlap window size. 
We note that for all cases, with and without noise and for varying overlap size, the modal amplitudes and frequencies are comparable to the ground truth with relatively small error, as shown in Fig~\ref{fig5}.

\begin{figure}
\vspace{-.125in}
  \centerline{\includegraphics[width=0.9\textwidth]{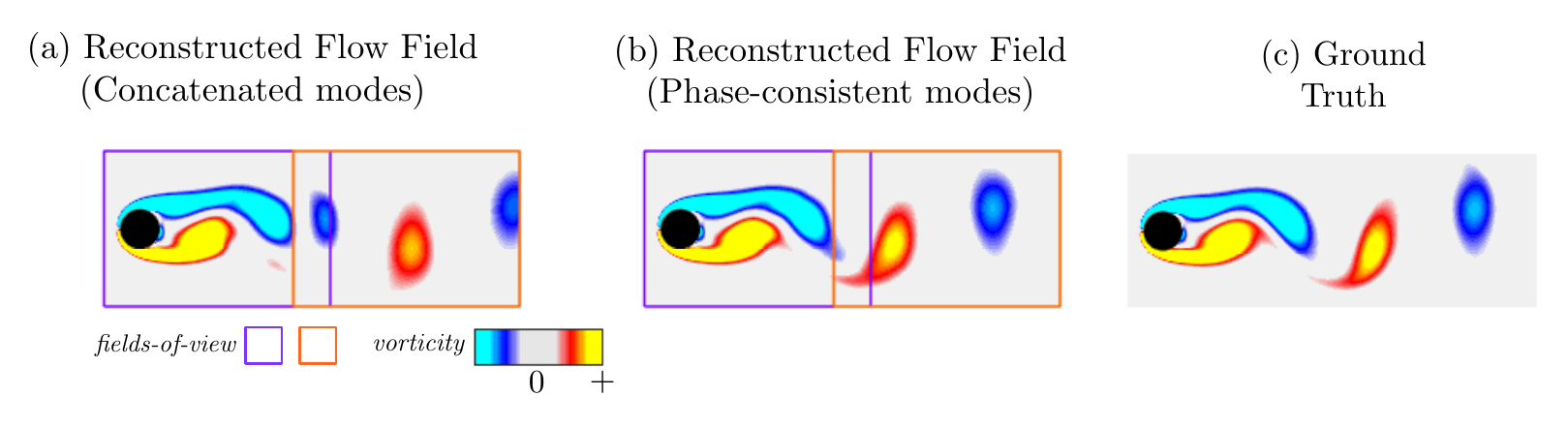}}
    \vspace{-.225in}
  \caption{Comparison between reconstructed flow fields from (a) concatenated modes, (b) phase-consistent DMD modes with ground truth flow field.}
    \vspace{-.125in}
\label{fig6}
\end{figure}

Finally, it is possible to reconstruct the phase-consistent flow field on the entire domain $\mathcal{D}$ from the phase-consistent DMD modes using Eq.~\eqref{eq2}. 
The flow field reconstruction from the concatenated modes, phase-consistent modes, and the ground truth flow field are shown in Fig.~\ref{fig6}.

\subsection{Mixing layer}
\label{subsec:mixing_layer}
Next, we demonstrate the phase-consistent DMD analysis with data from a two-dimensional, spatially developing, free shear layer flow. 
The shear layer is characterized by two initial inflow streams separated by a splitter plate, so that after the splitter plate ends the two flows interact with one another.  
The top stream has Mach number $M_1 = U_1/a_\infty = 0.4$ and the bottom stream has $M_2 = U_2/a_\infty = 0.1$. %
A free shear layer develops at the trailing edge of the plate, with mean velocity $\bar{U} = (U_1+U_2)/2$. 
The momentum thickness $\theta_0$ of the boundary layers of both streams is fixed at $10\%$ of the splitter plate thickness $w$, i.e., $\theta_0/w = 0.1$. 
The momentum-based Reynolds number is $Re_{\theta_0} = \rho_\infty \bar{U}\theta_0/\nu_\infty = 250$, where $\rho_\infty$ and $\nu_\infty$ are the free stream density and dynamic viscosity, respectively. 
The flow is obtained via direct numerical simulation using the compressible CharLES flow solver~\citep{bres2017unstructured}. 
The solver uses a second-order finite volume scheme and a third-order Runge-Kutta method for time integration. 
The domain is fixed so that $-15 \le x/w \le 400$ and $-200 \le y/w \le 200$, where $x$ and $y$ are the stream-wise and cross-stream directions, respectively. 
Additional simulation and validation details can be found in~\cite{yeh2017laminar}. 

\begin{figure}
  \centerline{\includegraphics[width=0.925\textwidth]{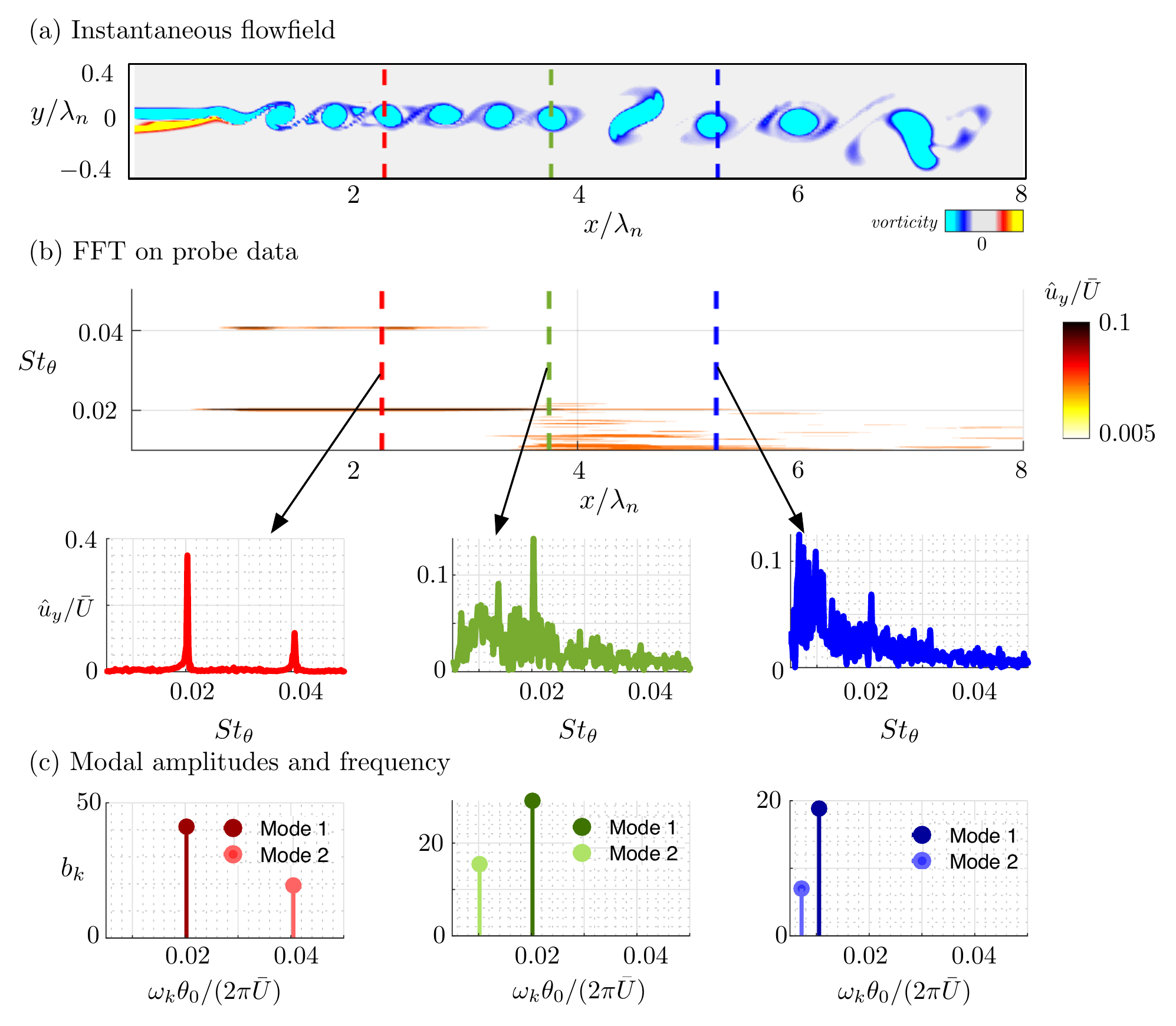}}
  \vspace{-.1in}
  \caption{Mixing layer flow at $Re_\theta = 250$: (a) Instantaneous vorticity field, (b) Spectral analysis of centerline vertical velocity at three streamwise stations, (c) DMD spectrum.}
\label{fig7}
\end{figure}

We define a fundamental shear-layer roll-up wavelength, $\lambda_n = \bar{U}/f_n$, where $f_n$ is the roll-up frequency.  
The dimensions of the computational domain are normalized by this wavelength. 
An instantaneous flow field is shown in Fig.~\ref{fig7}(a). 
The flow physics consist of four stages: (i) linear growth, corresponding to initial vortex roll-up due to Kelvin-Helmholtz instability, (ii) isolated vortex street consisting of compact vortices, (iii) nonlinear vortex pairing, and (iv) vortices deviating from the centerline. 
The spectral analysis of probes placed at streamwise stations $x/\lambda_n = 2.25, 3.75,$ and $5.25$ along the centerline ($y/\lambda_n = 0$) is shown in Fig.~\ref{fig7}(b). 
There is a strong peak corresponding to the roll-up frequency $St_\theta = f_n\theta_0/\bar{U} = 0.0203$ from $x/\lambda_n = 1$ to $3$. 
As the probe moves downstream, this frequency becomes weaker and subharmonic frequencies emerge. 
In general, the spectrum becomes increasingly broadband as the probe moves downstream. 

We test the phase-consistent DMD analysis in three sub-regions of the full flow field corresponding to different regimes of shear-layer flow physics; in each regime, we split the sub-region into two overlapping windows collected at different times to simulate two independent experiments.  
The full and overlapping domains for the three regimes are summarized in Table~\ref{table3}. 
The DMD modal amplitudes and frequencies are shown in Fig.~\ref{fig7}(c). 
The first regime is closest to the splitter plate and contains regions of linear growth and isolated vortex street; in this regime, the physics is predominantly linear, and DMD should be applicable.
The second regime is further downstream and contains nonlinear vortex pairing; the spectrum is broadband in this region, although there are still dominant peaks. 
The third regime is the furthest downstream and contains the vortex breakdown; the flow is not periodic or quasi-periodic in this region, and DMD is not applicable, so it expected that the phase-consistency analysis will fail.  
As in the cylinder flow example, data is available over the entire domain, so it is possible to compare our results with  DMD performed in the full domain for each regime.

\begin{table}[ht]
\centering 
\caption{Mixing layer analysis setup.}
\vspace{-.1in}
\begin{tabular}{c c c c c c c} 
\hline\hline 
 Full domain & $x/\lambda_n$ & $y/\lambda_n$ & Overlap. domains & $x/\lambda_n$ & $y/\lambda_n$ & $(\%)$ overlap \\ [0.5ex] 
\hline 
& & & $\mathcal{D}^{(1)}$ & $[0.5,2.5]$ & $[-0.75,0.75]$ & \\[-1ex] 
\raisebox{1.5ex}{$\mathcal{D}$(I)} 
& \raisebox{1.5ex}{$[0.5,4]$} & \raisebox{1.5ex}{$[-0.75,0.75]$} & $\mathcal{D}^{(2)}$ & $[2,4]$ & $[-0.75,0.75]$ & \raisebox{1.5ex}{14.3} \\[0.5ex] 
\hline 
& & & $\mathcal{D}^{(1)}$ & $[2,4]$ & $[-0.75,0.75]$ &\\[-1ex] 
\raisebox{1.5ex}{$\mathcal{D}$(II)} 
& \raisebox{1.5ex}{$[2,5.5]$} & \raisebox{1.5ex}{$[-0.75,0.75]$}  & $\mathcal{D}^{(2)}$ & $[3.5,5.5]$ & $[-0.75,0.75]$ & \raisebox{1.5ex}{14.3}  \\[0.5ex] 
\hline 
& & & $\mathcal{D}^{(1)}$ & $[3.5,5.5]$ & $[-0.75,0.75]$ & \\[-1ex] 
\raisebox{1.5ex}{$\mathcal{D}$(III)} 
& \raisebox{1.5ex}{$[3.5,7]$} & \raisebox{1.5ex}{$[-0.75,0.75]$}  & $\mathcal{D}^{(2)}$ & $[5,7]$ & $[-0.75,0.75]$ & \raisebox{1.5ex}{14.3} \\[0.5ex] 
\hline 
\end{tabular}
\label{table3} 
\vspace{-.1in}
\end{table}

\begin{figure}
  \centerline{\includegraphics[width=0.9\textwidth]{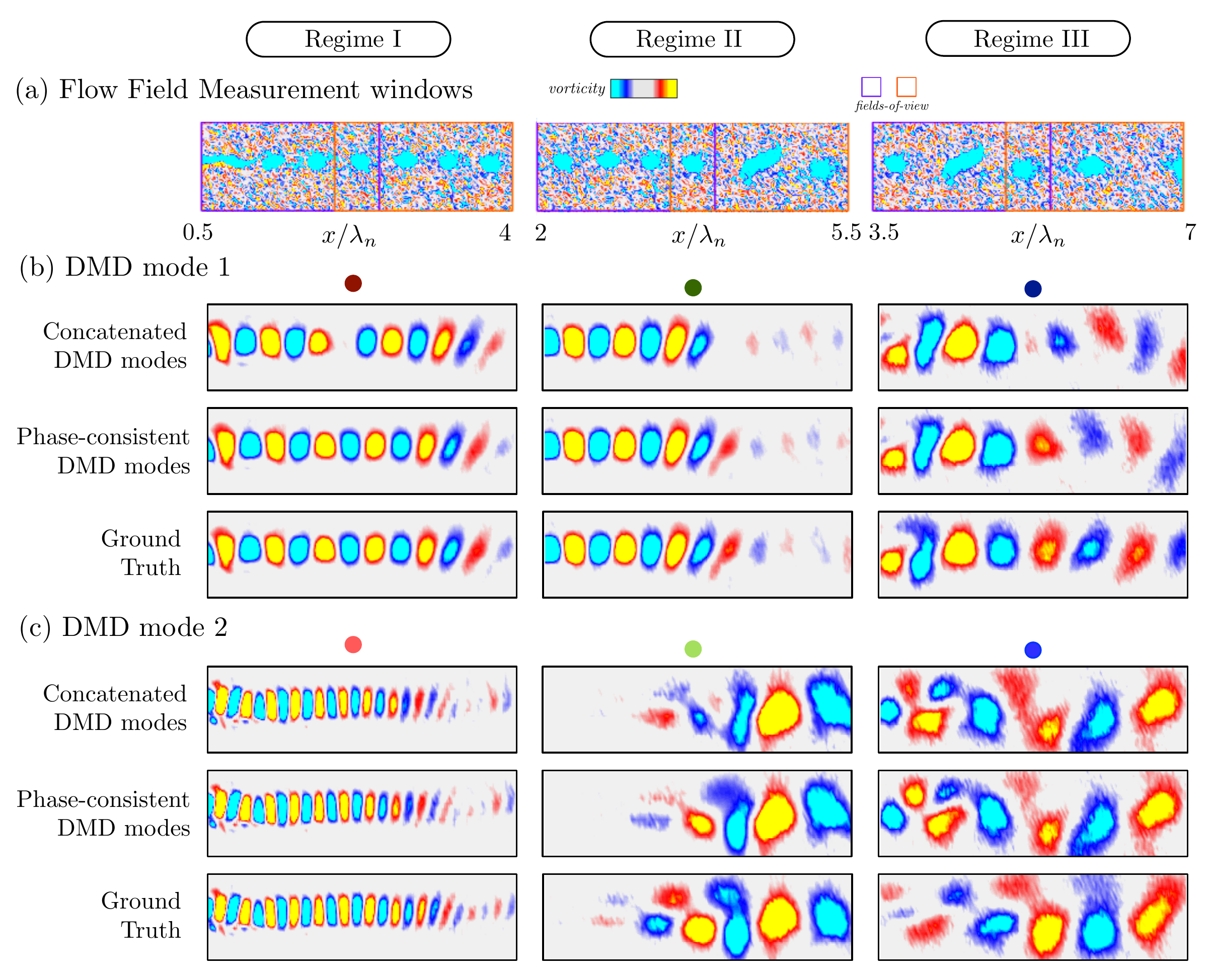}}
  \vspace{-.1in}
  \caption{Phase-consistency analysis of mixing layer flow with SNR $= 1$ at $Re_\theta = 250$: (a) Flow field snapshots collected over two overlapping fields-of-view. Comparison of concatenated modes, phase-consistent modes and ground truth for (b) DMD mode 1 and (c) DMD mode 2.}
\label{fig8}
  \vspace{-.15in}
\end{figure}

The phase-consistent DMD modes are shown for each regime in Fig.~\ref{fig8}.   
We have added white noise with SNR  $= 1$ to the flow field snapshots. 
 In regime I, the phase-consistent modes show close agreement with the ground truth modes, while the phase of the concatenated modes do not agree.  
 In regime II, there is a clear improvement in the phase-consistent modes over the concatenated modes.  
For this spatially developing flow, we see two characteristic spatially-dependent features in regime II: (i) the effect of the dominant mode behavior decaying and (ii) the inception of the sub-harmonic mode $2$. 
 In regime III, the dominant phase-consistent mode shows better agreement with ground truth, although the second mode doesn't show improvement over the concatenated mode.  
 The breakdown in the DMD modes in regime III is not attributed to the phase-consistency analysis but to the multi-domain DMD analysis, as DMD is not expected to perform well in such broadband, aperiodic flows.

 \begin{figure}
  \centerline{\includegraphics[width=0.9\textwidth]{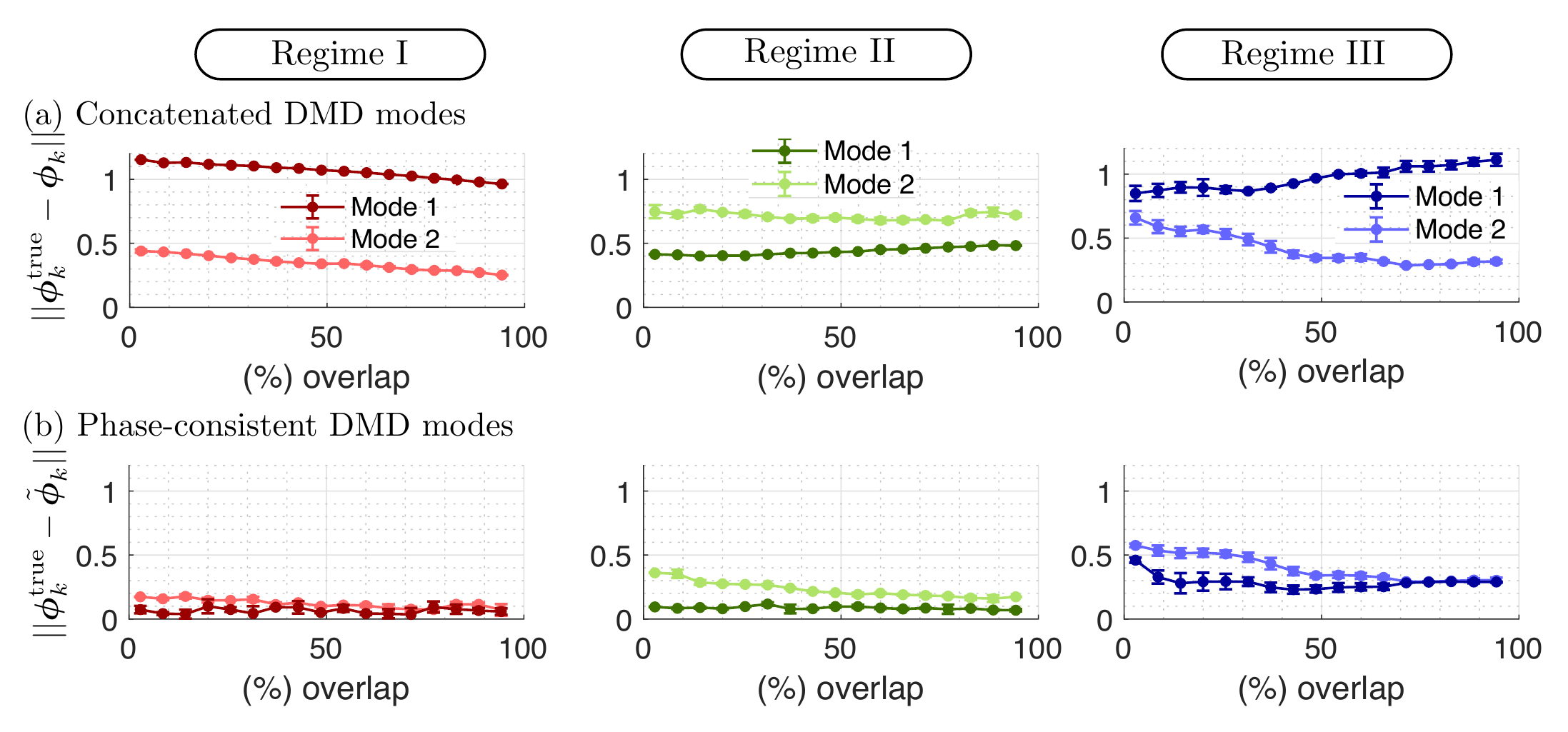}}
  \vspace{-.1in}
  \caption{Error in DMD modes for mixing layer flow with no noise and ${\text{SNR} = 1}$ (error bars) over three characteristic flow regimes.  (a) Error in concatenated DMD modes and (b) Error in phase-consistent DMD modes using approach B with overlap window size.}
\label{fig9}
  \vspace{-.15in}
\end{figure}

\begin{figure}
  \centerline{\includegraphics[width=0.9\textwidth]{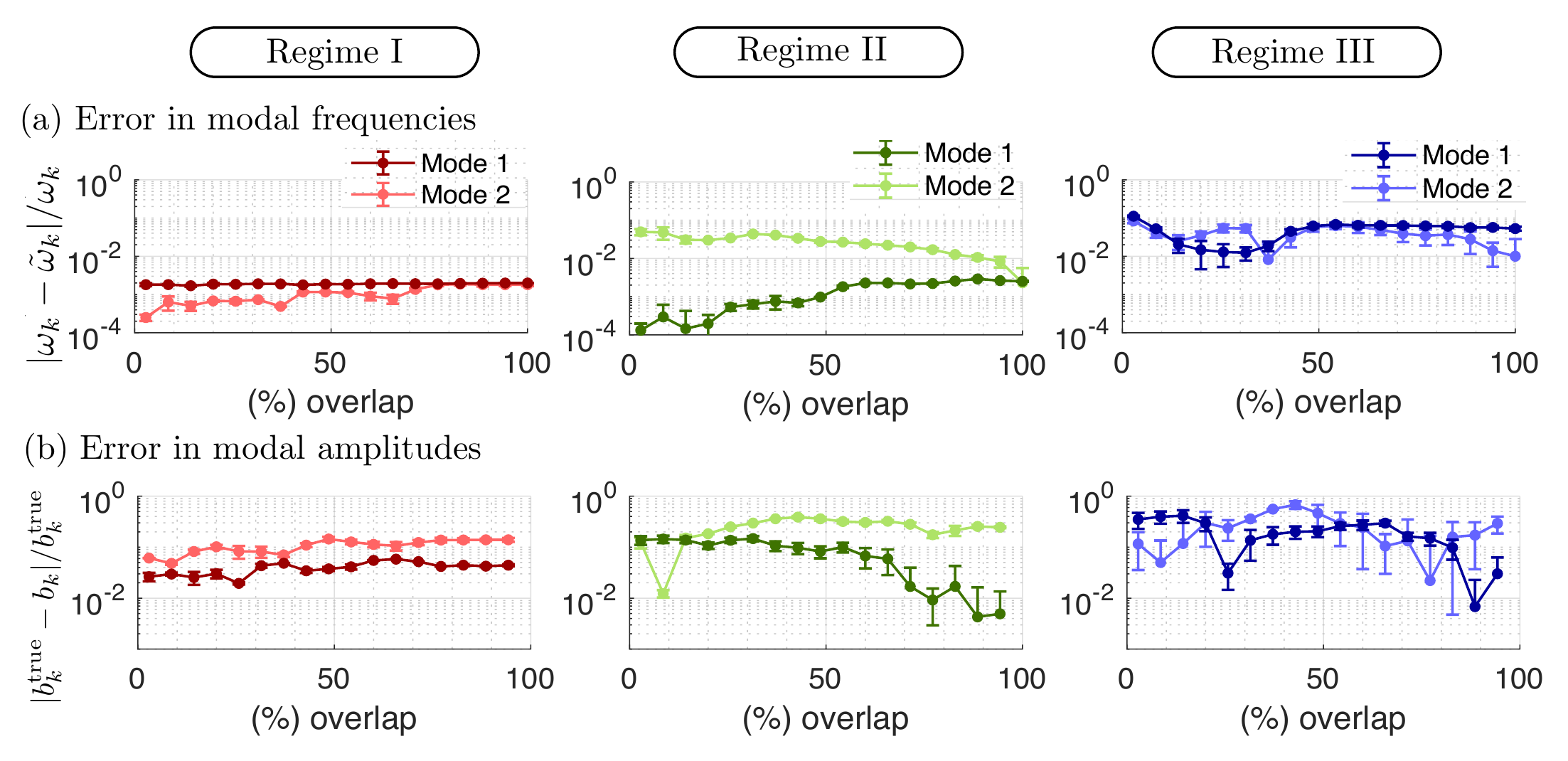}}
  \vspace{-.1in}
  \caption{Error in multi-domain DMD analysis for mixing layer flow with no noise and ${\text{SNR} = 1}$ (error bars) over three characteristic flow regimes: (a) Error in modal frequencies and (b) Error in modal amplitudes with overlap window size.}
\label{fig10}
  \vspace{-.15in}
\end{figure}

The error of concatenated DMD modes (${\bPhi}^{\text{cat}}$) and phase-consistent modes ($\tilde{\bPhi}^\cup$), compared with the ground truth modes  (${\bPhi}^{\cup}$), is shown for each regime in Fig.~\ref{fig9} for varying overlap size.  
The dots indicate the performance with no noise, and the error bars indicate the variability due to the addition of noise with SNR $= 1$ over $5$ realizations.  
The effect of noise appears to be minimal, as the variability is quite low in this example.  
The concatenated modes have significantly larger error than the phase-consistent modes in all cases. 
As the measurement window moves downstream from regime I to regime III, the error of the phase-consistent modes increases, which is to be expected as the periodic flow assumptions underlying DMD begin to break down.  
 We also evaluate the error in the modal frequencies and amplitudes from the multi-domain DMD analysis from   \S \ref{subsec:mdDMD} in Fig.~\ref{fig10}.  
 Overall, the error increases as the measurement window moves downstream, and the error is quite low for the upstream windows.

\subsection{Cross-flow turbine wake}
\label{subsec:cfturbine}
\begin{figure}
  \centerline{\includegraphics[width=.9\textwidth]{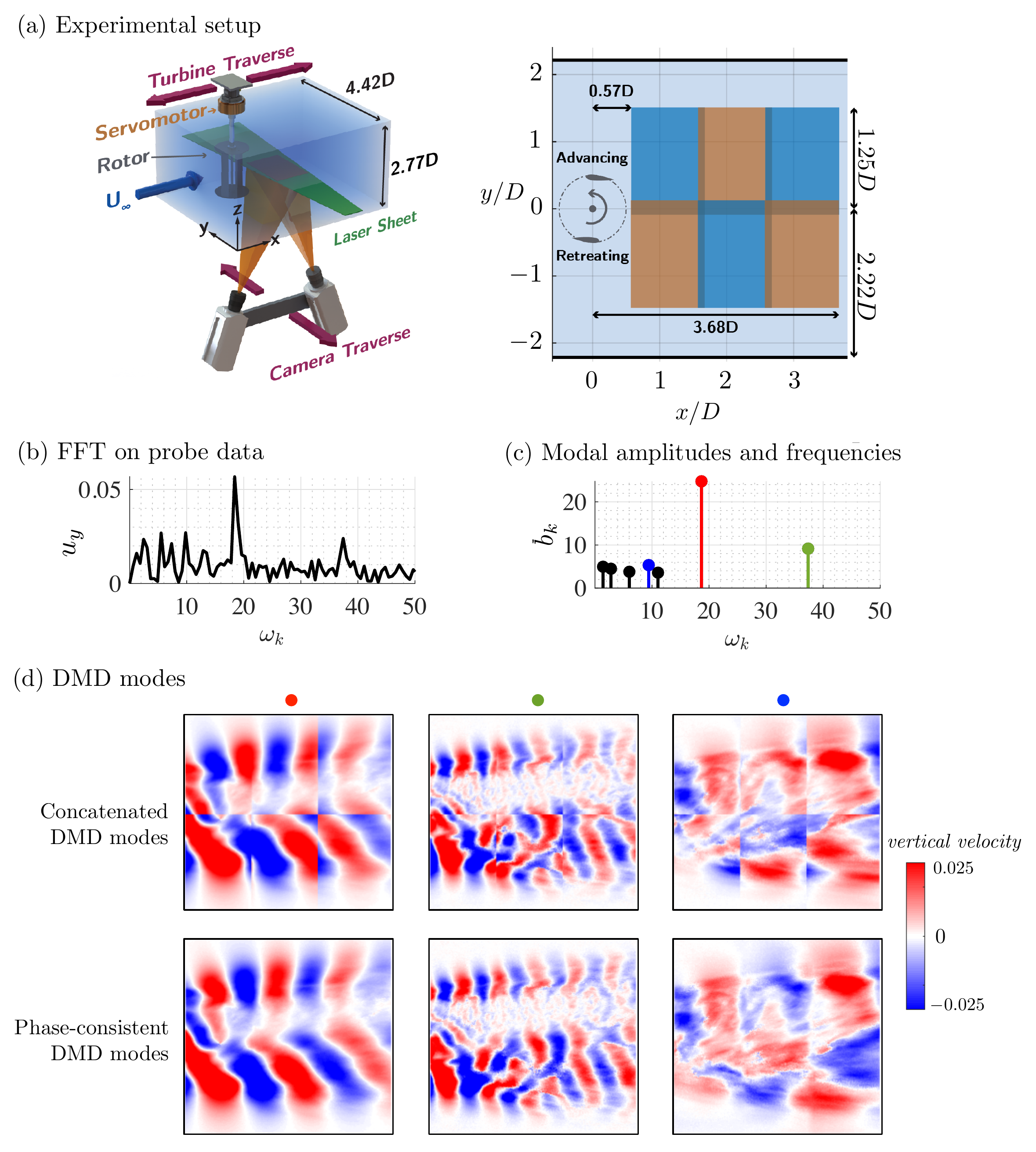}}
    \vspace{-.15in}
  \caption{Phase alignment of cross-flow turbine wake modes. (a) Experimental setup and fields of view, (b) Probe spectra, (c) DMD spectrum, and (d) comparison of concatenated and phase-consistent DMD modes.}
  \vspace{-.15in}
\label{fig11}
\end{figure}

The final demonstration of the phase-aligned DMD approach is for PIV data from six fields-of-view in the wake of a cross-flow turbine, shown in Fig.~\ref{fig11}(a). 
Here, we do not have ground truth data over the full domain to compare the results against, as in the previous two examples. 
Measurements in each of the six fields-of-view are taken independently using time-resolved stereo planar PIV; the dimensions of the field-of-view domains is summarized in Table~\ref{table4}. 
The turbine consists of two straight NACA0018 profile blades with chord length $c = 0.061\,$m and $c/R = 0.71$, where $R$ is the rotor radius. 
The turbine is operated under constant angular velocity at a tip-speed ratio of $\lambda = \Gamma R/U_\infty = 1.2$, where $\Gamma$ is the rotational velocity and $U_\infty = 0.7~$m/s is the freestream velocity. 
The rotor diameter-based Reynolds number is $Re_D = {D U_\infty}/{\nu} = 1.1\times 10^5$.

\begin{table}
\centering 
\caption{Cross-flow turbine analysis setup}
\vspace{-.1in}
\begin{tabular}{c c c c} 
\hline\hline 
Overlap. domains & $x/D$ & $y/D$ & $(\%)$ overlap \\ [0.5ex] 
\hline 
$\mathcal{D}^{(1)}$ & $[0.58,1.68]$ & $[-0.09,1.5]$ & \\
$\mathcal{D}^{(2)}$ & $[1.58,2.68]$ & $[-0.09,1.5]$ & \\
$\mathcal{D}^{(3)}$ & $[2.58,3.68]$ & $[-0.09,1.5]$ & \\[-1ex] 
$\mathcal{D}^{(4)}$ & $[0.58,1.68]$ & $[-1.47,0.12]$ & \raisebox{1.5ex}{16}\\
$\mathcal{D}^{(5)}$ & $[1.58,2.68]$ & $[-1.47,0.12]$ & \\ 
$\mathcal{D}^{(6)}$ & $[2.58,3.68]$ & $[-1.47,0.12]$ & \\[0.5ex] 
\hline 
\end{tabular}
\label{table4} 
\label{table}
\vspace{-.1in}
\end{table}

Data was collected at 100 Hz and was not locked to a specific blade position, making it difficult to align the data from the six fields-of-view. 
Measurements were taken at the mid-span of the turbine, in the plane normal to the axis of rotation. 
Illumination was provided by a Continuum TerraPIV Nd:YLF laser, and images were captured by two Phantom V641 cameras, with  $2560\times1600$  pixel resolution.
Cavitation bubbles from the flume recirculation pump were used as passive tracers and the velocity fields were calculated using iterative multi-grid processing. 
Measurement resolution was increased by capturing the wake using six overlapping fields-of-view, summarized in Table \ref{table4}. 
The combined measurement area starts 0.57$D$ downstream from the turbine axis, and extends 3.68$D$ downstream and 3$D$ in the cross-stream direction~\citep{strom2017intracycle}.
In the previous examples, only pairwise domains were considered for phase-consistency analysis; however, in this example, there are six overlapping domains, requiring the general phase-correction procedure in \S \ref{Sec:GeneralPhaseDMD}.

The vertical velocity probe spectra extracted at the center of the interrogation domains as well as the multi-domain DMD amplitudes and frequencies are shown in Fig.~\ref{fig11}(b) and (c), respectively. The first seven modes extracted by the multi-domain DMD algorithm contain the blade pass frequency and its first harmonic, followed by five lower-frequency modes, one of which appears to be a first sub-harmonic.
 The dominant three frequencies extracted are colored in Fig.~\ref{fig11}(c). We perform our phase-consistency analysis on the DMD modes corresponding to these frequencies, illustrated in Fig.~\ref{fig11}(d). The comparison between the concatenated modes (${\bPhi}^{\text{cat}}$) and phase-consistent modes ($\tilde{\bPhi}^\cup$) highlight the significance of the phase-alignment analysis proposed in this paper. The phase-consistent modes exhibit continuity in the modal structures, and shedding physics is consistent across the multiple overlapping windows.
  Due to this analysis, we can now gather additional insights into the underlying flow physics. 
  For example, the vortex shedding in the dominant mode corresponding to the blade pass frequency and its first harmonic both decay in the downstream direction, while large structures in the third mode, corresponding to the sub-harmonic frequency, increase in amplitude in the downstream direction.  
  This behavior suggests the transition from the near-wake shear layer structure to a bluff-body vortex shedding in the far wake, as observed by Araya et al~\cite{araya2017transition}.

\section{Conclusions}
\label{conc}

In this work, we have demonstrated a strategy to align the phases of DMD modes collected from multiple time series data in overlapping fields-of-view initialized at different starting times.  
This approach is ultimately motivated by experimental data collection, such as PIV measurements, where there is often a compromise between the spatial resolution and the size of measurement region. 
Thus, a large field-of-view with high resolution may require the composite of a number of datasets collected independently in overlapping small field-of-view windows.  
By enabling global flow reconstruction from data collected in multiple overlapping small field-of-view windows from different experimental runs, we can increase the size and resolution of flow field measurements while retaining the critical time evolution of the fluid dynamics.  

First, we show that simply concatenating the modes obtained by performing DMD in individual fields-of-view fails to yield globally phase-consistent modal structures.  
Next, we introduce a new technique to obtain phase-consistent DMD modes over multiple domains. 
This approach involves two steps.  First, DMD is computed on a large concatenated data set containing all of the overlapping domains simultaneously.  Second, an optimization procedure is used to align the phases of the resulting DMD modes from the multiple windows by minimizing mismatch in the overlap regions.  
For two overlapping spatial domains, we show that this phase optimization may be achieved by a simple regression procedure.  
For the case of multiple overlapping domains, we connect the more general optimization procedure to the invariance of DMD to unitary transformations, extending previous results~\cite{Brunton2015jcd}.  
The phase-consistency analysis is robust to noise and for varying sizes of the overlapping region.  
We also show that it is possible to obtain time-resolved composite flow fields over the entire global domain by reconstruction with the phase-consistent DMD modes.  

We have demonstrated this approach on several example systems from simulated and experimental fluid flows.  
In the numerical examples, we split the data into two overlapping domains and benchmark against ground-truth modes, providing a testbed to investigate the robustness of the method to noise and overlap. 
The first numerical example is the canonical flow past a circular cylinder, which is a strictly periodic flow.  
The second numerical example is the spatially developing mixing layer, which exhibits a spectrum that evolves and broadens as the measurement window moves downstream. 
Finally, the third example consists of experimental velocity fields obtained from PIV in six overlapping domains in the wake of a cross-flow turbine.  
In all examples, the phase-consistent DMD analysis yields accurate and consistent global modes and enables phase-aligned, composite reconstructions of the time-resolved flow field over the entire domain.  

There are a number of interesting avenues of future work that are suggested by this analysis. 
It will be important to continue to develop guidelines for when this phase-consistency approach will succeed or fail.  
Fundamentally, this involves a deeper understanding of when the standard DMD approach is expected to yield a sensible modal decomposition, for example how to interpret DMD for spatially evolving, non-stationary, and broadband flows.

\section*{Acknowledgements}  SLB gratefully acknowledges support from the Army Research Office (ARO W911NF-17-1-0306) and the Air Force Office of Scientific Research (AFOSR FA9550-18-1-0200).  
BWB and SLB gratefully acknowledge support from the Air Force Research Labs (AFRL FA8651-16-1-0003).  
BWB gratefully acknowledges support from the Air Force Office of Scientific Research (AFOSR FA9550-18-1-0114). 
  We are grateful to Brian Polagye, in whose flow facilities the PIV data was collected by BS.  We would also like to thank Isabel Scherl and Brian Polagye for valuable discussions.

\bibliographystyle{jfm}
 \begin{spacing}{.6}
 \small{
 \setlength{\bibsep}{5.pt}
 \bibliography{phaseDMD_references}
 }
 \end{spacing}

\end{document}